\documentclass[journal,a4paper]{IEEEtran}
\usepackage{cite}

\ifCLASSINFOpdf
   \usepackage[pdftex]{graphicx}
\else
\fi
\usepackage{calc}
\usepackage[cmex10]{amsmath}
\usepackage{bbm}
\usepackage{dsfont}
\usepackage{amsfonts,latexsym,amsmath}
\usepackage{epstopdf}
\usepackage{amssymb}
\usepackage{amsthm}

\usepackage{tikz}
\usetikzlibrary{positioning,shadows,arrows}
\usepackage{caption}
\usepackage{algorithmic}

\usepackage{array}

\usepackage{bm}
\usepackage{mdwmath}
\usepackage{mdwtab}

\usepackage{eqparbox}
\newcommand{\RR}{\mathbb{R}}

\DeclareMathOperator{\prob}{\mathbb{P}}

\def\vec#1{{\bf #1}}

\def\bH{{\bf H}}

\def\b0{{\rm 0}}

\newfont{\bb}{msbm10 scaled 1100}

 \usepackage{pgfplots}
 
 \usepackage{bm}
\usepackage{subfigure,tikz}
 \usepackage{graphicx}
  \usepackage{color} 
  \usepackage{transparent}
  \usepackage{amsthm}

\usetikzlibrary{arrows,shapes,decorations,backgrounds}

\usepackage{import}
\DeclareMathOperator{\galrho}{\rho}
\DeclareMathOperator{\s2}{\sigma^2}

\newtheorem{theorem}{Theorem}
\newtheorem{definition}{Definition}
\newtheorem{remark}{Remark}
\newtheorem{cor}{Corollary}

\begin{document}

\title{Gallager Bound for MIMO Channels:\\ Large-$N$ Asymptotics}

\author{Apostolos~Karadimitrakis$^{1}$, Aris~L.~Moustakas$^{1}$ and Romain Couillet$^2$
\thanks{(1): Department of Physics, National and Kapodistrian University of Athens, Greece}
\thanks{(2): Centrale-Sup\'elec, Paris, France}
\thanks{R. Couillet acknowledges partial support from HUAWEI RMTin5G and ERC MORE 305.123.}}

\maketitle

\begin{abstract}
\boldmath
The use of multiple antenna arrays in transmission and reception has become an integral part of modern wireless communications. To quantify the performance of such systems, the evaluation of bounds on the error probability of realistic finite length codewords is important. In this paper, we analyze the standard Gallager error bound for both constraints of maximum average power and maximum instantaneous power. Applying techniques from random matrix theory, we obtain analytic expressions of the error exponent when the length of the codeword increases to infinity at a fixed ratio with the antenna array dimensions. Analyzing its behavior at rates close to the ergodic rate, we find that the Gallager error bound becomes asymptotically close to an upper error bound obtained recently  by Hoydis et al. 2015.  We also obtain an expression for the Gallager exponent in the case when the codelength spans several Rayleigh fading blocks, hence taking into account the situation when the channel varies during each transmission.
\end{abstract}

\begin{IEEEkeywords}
Error bound, Gallager, Large Deviation, MIMO, Wireless Communication 
\end{IEEEkeywords}

\IEEEpeerreviewmaketitle

\section{Introduction}

\IEEEPARstart{I}{}n recent years wireless communications have experienced an unprecedented growth in the number of users, data rate and throughput volumes. To meet this demand, MIMO techniques, at both the link and network level, promise to dramatically increase the information throughput. For fading channels, the standard metric to characterize the performance of the link is the outage capacity \cite{ozarow1994information}, which corresponds to the throughput for a fixed outage probability. However, the outage capacity corresponds to infinitely long codewords. To deal with the realistic case of finite length codewords, Gallager \cite{gallager1968information} proposed a simple yet effective bound to the probability of error, as a function of rate and codeword length $T$. In its original version, as well as in more recent variations \cite{Shamai2003_VariationsGallager} this bound focused on single antenna links. There has been a number of extensions of the Gallager bound. For example, in \cite{shin2009gallager} the Gallager's random coding error exponent was derived for MIMO Rayleigh block-fading channels, however, the expressions, while valid for all antenna sizes, are quite cumbersome to compute and analyze for any reasonably sized antenna array. In \cite{karagiannidis2013gallager, xue2012performance} expressions for Gallager's exponent were derived for space-time-block-coding (STBC) MIMO channels for non-Rayleigh fading models. However, STBC reception effectively corresponds to a single antenna link with increased diversity. 

More recently, optimal bounds of the error probability for large but finite codewords have been established for single-link communications\cite{hayashi2009information, polyanskiy2010channel}. These results are of a central--limit--theoretic nature, in that they are valid for large blocklengths with the rate converging to the ergodic rate at a fixed error probability.  Similar results were obtained  for MIMO systems in \cite{hoydis2015second}, where the number of antennas also goes to infinity at a fixed ratio with $T$. In contrast to the Gallager bound this approach does not capture the tails of the error probability, i.e. when the rate deviation per antenna from the ergodic rate is finite. 

In this paper we apply random matrix theory to evaluate the error probability exponent of the Gallager bound when  the blocklength $T$, the number of $N$ transmitting, and $K$ receiving antennas, and the rate $R$ all become large but at fixed ratios $\alpha=T/N$, $\beta=K/N$, $r=R/N$. Our large deviation result is valid for all normalized rates $0<r<r_{erg}$. When we evaluate the error exponent for small $|r_{erg}-r|\ll 1$, our results match with the upper bound obtained by \cite{hoydis2015second}. While the asymptotic limit of large antenna numbers is somewhat idealized, it is known from other works, e.g. \cite{kazakopoulos2011living} that even for moderate antenna numbers the asymptotic results become quite accurate.
In addition, we explore the impact of fading in the channel by allowing the channel to take $Q$ independent realizations within a codeword of length $T$. Our approach, which maps the random matrix problem to a gas of Coulomb charges on a line, was first introduced in the context of statistical physics by Dyson \cite{dyson1962statistical} and recently in \cite{kazakopoulos2011living} and \cite{karadimitrakis2014outage} in the context of information theory and communications.

\subsection{Outline and Notations}
In the next section we formulate the problem and present the main results. In Section III we discuss our findings in representative limits, while in Section IV we conclude. 

We use upper case letters in bold font to denote matrices, $ \mbox{e.g.},~\mathbf{X}$, with entries given by $X_{ab}$. The superscript $\dagger$ denotes the Hermitian transpose operation and $\mathbf{I}_N$ represents the $N$-dimensional identity matrix.

\section{Problem Formulation and Results}
\subsection{Channel Model and Capacity}
\label{sec:channel_model_cap}
Let us consider a MIMO link with $N$ transmit and $K$ receive antennas and analyze the transmission of $T$ symbols. We assume a block fading channel, which  remains constant over $\tau _Q =\left[\frac{T}{Q}\right]$ symbols and changes independently after each such coherence time \cite{berry2002communication}. Hence $\tau_Q$ is a parameter indicated by the bandwidth of the system and the fading statistics of the channel. Therefore, the memoryless channel reads
\begin{eqnarray}
\mathbf{Y}_q = \mathbf{H}_q\mathbf{X}_q + \sigma \mathbf{W}_q
\end{eqnarray}
for $q = 1\dots Q$, where $\mathbf{Y}_q \in \mathbb{C}^{K \times \tau _Q}$ is the received signal matrix during the $q$th block, $\mathbf{H}_q~ \in~ \mathbb{C}^{K\times N}$ is the channel matrix, whose entries are independent and identically distributed (i.i.d.) $\mathcal{CN}(0,\frac{1}{N})$, $\mathbf{X}_q \in \mathbb{C}^{N\times \tau _Q}$ is the transmitted signal matrix and $\sigma \mathbf{W}_q\in \mathbb{C}^{K \times \tau _Q}$ is the noise matrix  with entries i.i.d. following $\mathcal{CN}(0,\sigma ^2)$. For notational convenience we will denote 
$ \mathbf{Y}=[\mathbf{Y}_1,\ldots,\mathbf{Y}_Q]$, $\mathbf{X}=[\mathbf{X}_1,\ldots,\mathbf{X}_Q]$, etc. The transmitter has only statistical knowledge of the channel, while the receiver knows it perfectly e.g., using a pilot signal. The mutual information per channel use over the $q$th block for Gaussian input with i.i.d. entries following ${\mathcal CN}(0,1)$ is given by
\begin{align}
C_{q} (\sigma ^2,\mathbf{H}_q) =   \log \det \left( \mathbf{I}_N + \frac{1}{\s2}\mathbf{H}_q\mathbf{H}_q ^\dagger  \right).
\end{align}
The joint distribution of eigenvalues of $\mathbf{H}_q\mathbf{H}_q^\dagger$ is
\begin{align}
\label{Eq:JPDF}
P_\mathbf{\pmb{\lambda}}(\lambda_1 \dots \lambda_N) &= \frac{1}{\mathcal{Z}_N} \prod_{N<i<j\leq K}|\lambda _i - \lambda_j|^2
 \prod _i w(\lambda_i)\nonumber\\
& = \frac{1}{\mathcal{Z}_N} e^{-N^2E(\pmb{\lambda})}  ,
\end{align}
where $\mathcal{Z}_N$ is the normalization constant, $w(\lambda)$ is a weight function, which depends on the statistics of $\bH_q$ and the exponent $E(\pmb{\lambda})$ is an energy functional of the eigenvalues $\{ \lambda _i \}$ that will become useful later. Fot the case of complex Gaussian channels, which is the focus of this paper, the form of  the weight function is $w(x)=x^{K-N}e^{-Nx}$. There are a number of other random matrix models for which the joint distribution of eigenvalues takes the same form with different  realizations of $w(x)$ which we will briefly comment later in the paper. The value of the mutual information per antenna $C_q(\s2,\mathbf{H}_q)/N$ converges weakly to a deterministic value in the large $N$ limit, given by the ergodic average of the mutual information \cite{moustakas2003mimo} (Eq. 105-106),
\begin{align}
r_{erg}(\beta,\s2) = \log u + \beta \log\left[ 1+ \frac{1}{u\s2}\right] - (1 - u^{-1}),
\end{align}
with
\begin{align}
\label{Eq:u}
u = \frac{1}{2\s2}\left( \s2 + \beta -1 + \sqrt{(\s2 +(\beta-1))^2 +4\s2} \right),
\end{align}
where $\beta = \frac{K}{N} >1$. The empirical eigenvalue density of $\mathbf{H}_q\mathbf{H}_q^\dagger$ converges weakly to the well-known Mar\v{c}enko-Pastur distribution \cite{tulino2004random} (Equation 1.12) 
\begin{align}\label{Eq:MarcPastur}
p_{0}(x) =  \left\lbrace
                 \begin{array}{ll}
                  \frac{\sqrt{(b_0-x)(x-a_0)}}{2\pi x}, ~\mbox{for}~x \in [a_0,b_0]\\
                  0, ~~ \mbox{otherwise}.                 
                \end{array}
                \right.
\end{align}
where $a_{0},b_{0} = (\sqrt{\beta}\pm 1)^2$ are the endpoints of its support.

In the infinite codelength limit, the effect of the channel fading is captured through the optimal outage error probability \cite{ozarow1994information} over the channel matrix $\mathbf{H}_q$, given by $p_{out}=\prob(C/N<r)$ (in the case of $Q=1$, $C\equiv C_{1}$). The exponent of the outage probability was analyzed in \cite{kazakopoulos2011living} when the number of antennas becomes large.  There it was shown that when $K,N\to\infty$ with $\beta=K/N$ fixed, the outage probability behaves as
\begin{align}
\label{Eq:E_beta_r}
\lim_{N\to \infty} \frac{1}{N^2}\log\prob\left(\frac{C}{N}<r\right)=
-E_{out}(r),
\end{align}
where $E_{out}(r)$ close to $r=r_{erg}$ behaves as
\begin{align}
  \label{Eq:Gauss_approx_outage}
  E_{out}(r) = \frac{(r-r_{erg})^2}{2 v_{\infty}} + o\left( (r-r_{erg})^2 \right),
  \end{align}
where
\begin{align}\label{Eq:vopt_def}
v_{\infty} = -\log\left(1 - \frac{(1-u)^2}{\beta u^2}\right).
\end{align}
The above quantity is the dispersion of the mutual information distribution in the infinite codelength limit and will be called hereafter infinite codelength dispersion, in accordance with the names used for similar quantities in \cite{polyanskiy2010channel,altug2014moderate,hoydis2015second}.

\subsection{Gallager Exponent for Power-Constrained Input Alphabets}
On the other hand, for finite codelength $T$, one can estimate the error probability by using the so-called Gallager bound. Specifically, the error probability of transmission at a code rate of $R=Nr$ for a given instantiation of $\{\mathbf{H}_q \}$, $\mathbb{P}(\mathcal{E}|\{\mathbf{H}_q \})$ of a discrete memoryless channel without feedback and maximum likelihood (ML) decoding is bounded by (see Eq. (7.3.20) in \cite{gallager1968information})
\begin{align}
\label{Eq:E}
\prob(\mathcal{E}|\{\mathbf{H}_q\}) \leq &\nonumber\\ 
 e^{TNr}\int & d\vec{Y} \left[ \int d\vec{X} ~ \mu_{con}(\mathbf{X})\left[\mu (\mathbf{Y}|\mathbf{X},\{\mathbf{H}_q\})\right]^{\frac{1}{1+\galrho}}\right]^{1+\galrho},
\end{align}
where $\galrho\in [0,1]$, $\mu (\mathbf{Y}|\mathbf{X},\{\mathbf{H}_q\})$ is the distribution of the noise $\sigma \mathbf{W}$, while $\mu_{con}(\mathbf{X})$ is the distribution of $\mathbf{X}$ constrained to inputs such that only codewords with
\begin{align}
\label{Eq:Energy_Contraint}
\mbox{Tr}\left[\mathbf{X}^\dagger \mathbf{X}\right] \leq NT
\end{align}
are used. This constraint can be enforced as an inequality by following (Eq. (7.3.17)) in\cite{gallager1968information}  to observe that 
\begin{align}\label{Eq:mu_constrained_def}
\mu_{con}(\mathbf{X})\leq \bar{c}~\mu(\mathbf{X}) e^{s\left(\mbox{Tr}\left[\mathbf{X}^\dagger\mathbf{X}\right]-NT\right)},
\end{align}
for any $s>0$, where $\mu(\mathbf{X})$ is the unconstrained input distribution assumed henceforth to be Gaussian and $ \bar{c}$ a normalization constant. 
Integrating over $\mathbf{X},~\mathbf{Y}$ we obtain
\begin{align}
\label{Eq:E(RH)}
\log\prob(\mathcal{E}|\{\mathbf{H}_q\}) \leq &\nonumber\\
  -\frac{T}{Q}\sum_{q=1}^{Q}&\bigg[
\galrho\log\mbox{det} \left( 1+ \frac{1}{(1+\galrho)(1-s)\s2}\mathbf{H}_q\mathbf{H}_q^\dagger \right)\nonumber\\
 -\galrho rN& +(1+\galrho)N\left(s+\log(1-s)\right)\bigg],
\end{align}
after omitting the normalization term $(\galrho+1)\log \bar{c}$, which can be shown to be subleading in $N$ \cite{gallager1968information}. After averaging $\prob(\mathcal{E}|\{\mathbf{H}_q\})$ over $\{\mathbf{H}_q\}$ and optimizing over the values of $\galrho,~s$, we find that $\prob(\mathcal{E})$, the average error rate after jointly decoding the total message sent over $Q$ blocks is bounded by 
\begin{align}
\mathbb{P}(\mathcal{E}) &= \mathbb{E}_{\{\mathbf{H}_q\}}\left[ \mathbb{P}\left(\mathcal{E}|\{\mathbf{H}_q\}\right) \right] \leq \mathbb{E}_{\{\mathbf{H}_q\}} \left[ e^{-N^2E(r|\{\mathbf{H}_q\})} \right],
\label{Eq:EN_def0}
\end{align}
where 
\begin{align}
\label{Eq:E(RH)_exp}
E(r|\{\mathbf{H}_q\})=&\nonumber\\
 \max_{\substack{\galrho \in [0,1]\\s\in [0,1)}} \bigg\lbrace \frac{\alpha}{Q}\sum_{q=1}^{Q}&\bigg[
\frac{\galrho}{N}\log\mbox{det} \left( 1+ \frac{1}{(1+\galrho)(1-s)\s2}\mathbf{H}_q\mathbf{H}_q^\dagger \right)\nonumber\\
 -&\galrho r +(1+\galrho)\left(s+\log(1-s)\right)\bigg] \bigg\rbrace .
\end{align}
In the above, $\alpha=T/N$ and $r=R/N$ is the per-antenna rate and $\sigma^{-2}$ is the SNR. We then define the Gallager exponent as
\begin{align}
E_N(r) = -\frac{1}{N^2} \log \mathbb{E}_{\{\mathbf{H}_q\}} \left[ e^{-N^2E(r|\{\mathbf{H}_q\})} \right].
\label{Eq:EN_def}
\end{align}

It should be stressed that while in single link transmission schemes the exponent of the probability of error scales with the blocklength $T$, in MIMO systems it should be proportional to $NT$, which is the number of symbols transmitted. To be able to compare with the infinite codelength error exponent defined in the previous section, we have chosen to re-scale the error exponent in the same way (i.e. with $N^2$), adding a factor of $\alpha$ in \eqref{Eq:E(RH)}.
We then take the limit $N, K, T\to\infty$, while at the same time keeping the ratios $\beta=K/N$ and $\alpha=T/N$ fixed.  The analytic evaluation of the error exponent $ E_N(r)$ in this limit is the main result of this paper and is summarized by the following theorem.

\begin{theorem}
\label{Th:E_beta_gr_1}
The limit of the error exponent ${E(r) = \lim _{N\rightarrow \infty} E_N(r)}$ exists and can be expressed as
\begin{align}
\label{Eq:E(r)_gen}
E(r) &= Q\max_{\substack{\galrho \in [0,1]\\s\in [0,1)}}\bigg[-\int_a^b\int_a^b \log|x-y|p^*(x) p(y)dxdy\nonumber\\
& + \int_a^b\left( x-(\beta-1)\right)\log(x)p^*(x)dx\nonumber\\
&+\frac{\alpha}{Q}\bigg(\galrho\int_a^b\log\left(1 + \frac{x}{z_{\galrho s}} \right)p^*(x)dx\nonumber\\
&- \galrho r +(1+\galrho)\left(s+\log(1-s)\right)\bigg) \nonumber \\
&-\frac{1}{2}\left( 3\beta -\beta ^2\log \beta +(\beta-1)^2\log(\beta-1)\right)\bigg], 
\end{align}
where
\begin{align}\label{eq:mu_x_2}
  p^*(x) =& \frac{\sqrt{(x-a)(b-x)}}{2\pi x(x+z_{\galrho s})}\nonumber\\
   &\times\left[x+z_{\galrho s} +\frac{\alpha \galrho z_{\galrho s}}{Q\sqrt{(z_{\galrho s}+a)(z_{\galrho s}+b)}}\right],
\end{align}
and $z_{\galrho s} = (1+ \galrho)(1-s)\s2$. 
The values of the parameters $a$, $b$ and $s$, as functions of $\galrho$, are the unique solutions of the following equations: 
\begin{align}
\label{eq:mu(a)=0}
\frac{\beta -1}{\sqrt{ab}} - \frac{\galrho \alpha}{Q\sqrt{(z_{\galrho s}+a)(z_{\galrho s}+b)}} =1,
\end{align}
\begin{align}\label{eq:mu_norm_condition}
  a+b+2\frac{\galrho\alpha}{Q} -2(\beta + 1) = \frac{2\galrho\alpha z_{\galrho s}}{Q\sqrt{(a+z_{\galrho s})(b+z_{\galrho s})}},
\end{align}
\begin{align}\label{eq:s_condition}
 s = \frac{\galrho}{4(1+\galrho)} \left(\sqrt{z_{\galrho s}+b}-\sqrt{z_{\galrho s}+a}\right)^2.
\end{align}
Having determined these parameters as functions of $\galrho$, $\galrho$ is determined from $r$ as follows. Defining the function $\bar{r}(\galrho)$ as
\begin{align} 
\label{Eq:r_bar}
 \bar{r}(\galrho)&= \log(1-s)+\int_a^b p^*(x) \log\left(1+\frac{x}{z_{\galrho s}}\right)dx \\ 
 \label{eq:mut_info_beta>1}
& =  \log\frac{\Delta (1-s)}{z_{\galrho s}}\nonumber\\
& + \frac{\Delta}{2}\left(1+\frac{\galrho \alpha}{\sqrt{(z_{\galrho s}+a)(z_{\galrho s}+b)}}\right) G\left( \frac{z_{\galrho s}+a}{\Delta},\frac{a}{\Delta} \right)\nonumber\\
&-\frac{\Delta\galrho \alpha}{2\sqrt{(z_{\galrho s}+a)(z_{\galrho s}+b)}}
G\left( \frac{z_{\galrho s}+a}{\Delta},\frac{z_{\galrho s}+a}{\Delta} \right)
\end{align}
where the function $G(x,y)$ the following known integral \cite{Chen1996_EigDistributionsLaguerre}  
\begin{align}
G(x,y)	=& \frac{1}{\pi}\int_0^1 \sqrt{t(1-t)} \frac{\log(t+x)}{t+y}\:d t\nonumber \\ 
		=& -2\sqrt{y(1+y)}\log\left[\frac{\sqrt{x(1+y)}+\sqrt{y(1+x)}}{\sqrt{1+y}+\sqrt{y}}\right]\nonumber \\ 
  +& \left(1+2y\right)\log\left[\frac{\sqrt{1+x}+\sqrt{x}}{2}\right]- \frac{1}{2}\left(\sqrt{1+x}-\sqrt{x}\right)^2,
\end{align}
and setting $r_1=\bar{r}(1)$ we have 
\begin{align}
\galrho(r) = \left\{
\begin{tabular}{ c c }
 $1$ & $r\leq r_1$\\ 
 ${\bar r}^{-1}(r)$ & $r>r_1$     
\end{tabular}
\right.
\end{align}
where $\bar{r}^{-1}$ indicates the inverse function of $\bar{r}$. 
\end{theorem}
The proof of Theorem~\ref{Th:E_beta_gr_1} can be found in Appendix~\ref{Appendix:Proof1}.
\footnote{$a$ and $b$ are the endpoints of the support of $p^*(x)$ and should not be confused with $\alpha=T/N$ and $\beta=K/N$.}

\begin{remark}
 \normalfont 
$p^*(x)$ defined in \eqref{eq:mu_x_2} and appearing in \eqref{Eq:E(r)_gen} and \eqref{Eq:r_bar} can be interpreted as a density of eigenvalues and exhibits a square root singularity at the limits of its support, just as the Mar\v{c}enko -- Pastur density \cite{PasturEigDens}. From physical point of view, $p^* (x) $ corresponds to the equilibrium charge density in the Coulomb gas picture, when the energy function is given by $E(r)$. From a practical point of view, it corresponds to the empirical distribution of observed eigenvalues $\{ \lambda _i \}$ of the realized channel matrices, which balance the occurrence probability of such channel matrices with the corresponding coding error probability, when operating at a given normalized rate $r$, $\alpha$ and $\beta$. 
\end{remark}
\begin{remark}
 \normalfont 
Setting $s=0$ in \eqref{Eq:mu_constrained_def} corresponds to an unconstrained Gaussian input distribution. Hence,  the corresponding solution of \eqref{Eq:E(r)_gen} will be  the Gallager exponent for unconstrained Gaussian inputs, which is expected to be smaller.
\end{remark}
\begin{remark}\label{rem:E(r,Q)}
 \normalfont 
From the equations of the above theorem we immediately see that the $Q$-dependence of $E(r)$ has the following form: $E(r,\alpha,Q)=QE(r,\frac{\alpha}{Q},1)$, where we explicitly included the dependence of $E(r)$ on $\alpha$ and $Q$. This allows us to make all calculations for $Q=1$ and in the end to re-scale $E(r)$ and $\alpha$ accordingly. 
\end{remark}

\begin{cor}
For $\beta > 1$ the above expression for the error exponent can be calculated in closed form to read  
\begin{align}
\label{Eq:ClosedForm_E(r)}
 E(r) &=
 Q\Bigg[\frac{\Delta ^2}{32} -\frac{\alpha\galrho r}{Q} + \frac{a}{2} -\log \Delta - \frac{\beta -1}{2}\log(a\Delta)\nonumber\\
  +&\frac{\alpha(1+\galrho)}{Q}\left(s+\log(1-s)\right)  \\
 +&\frac{ \alpha\galrho}{2Q}\left( \log(1 + a/z_{\galrho s}) + z_{\galrho s}\frac{\left( \sqrt{z_{\galrho s} + b} - \sqrt{z_{\galrho s}+a} \right)^2}{4\sqrt{(z_{\galrho s}+a)(z_{\galrho s}+b)}} \right)\nonumber\\
 +&\frac{\Delta \alpha\galrho}{2Q\sqrt{(z_{\galrho s}+a)(z_{\galrho s}+b)}}\nonumber\\
  &\times\left[ G\left( 0, \frac{z_{\galrho s}+a}{\Delta} \right) +\frac{\beta -1}{2}G\left( \frac{a}{\Delta},\frac{z_{\galrho s}+a}{\Delta} \right) \right]\nonumber\\
 -& \frac{\Delta}{2}\left( 1 + \frac{\alpha \galrho}{Q\sqrt{(z_{\galrho s}+a)(z_{\galrho s}+b)}} \right)\nonumber\\
 &\times\left[ G\left( 0, \frac{a}{\Delta} \right) + \frac{\beta -1}{2}G\left(\frac{a}{\Delta},\frac{a}{\Delta} \right) \right]\nonumber\\
 -&\frac{1}{2}\left( 3\beta -\beta ^2\log \beta +(\beta-1)^2\log(\beta-1)\right)\nonumber\\
 +&\frac{\alpha \galrho}{2Q} \Bigg[\log\left(\frac{\Delta}{z_{\galrho s}}\right)\nonumber\\ 
  -&\frac{\Delta\alpha \galrho}{2Q\sqrt{(z_{\galrho s}+a)(z_{\galrho s}+b)}} G\left(\frac{z_{\galrho s}+a}{\Delta},\frac{z_{\galrho s}+a}{\Delta} \right)
 \nonumber\\ 
+&\left(\frac{\Delta}{2}+\frac{\alpha\galrho \Delta}{2Q\sqrt{(z_{\galrho s}+a)(z_{\galrho s}+b)}}\right) G\left(\frac{z_{\galrho s}+a}{\Delta},\frac{a}{\Delta} \right)\Bigg]\nonumber
\end{align}
where $\Delta=b-a$.
\end{cor}
\begin{cor}
 In the special case $\beta =1$ the lower limit of the support of $p(x)$ becomes zero, i.e. $a=0$. In this case \eqref{eq:mu(a)=0} (which results from the continuity condition $p(a)=0$) does not hold. However, we can obtain $E(r)$ by setting $a=0$, $\beta=1$ in equations \eqref{eq:mu_norm_condition}, \eqref{Eq:r_bar},  \eqref{Eq:ClosedForm_E(r)}. Then $E(r)$ reads
 \begin{align}
 E(r) =&  Q\Bigg[\frac{\alpha\galrho}{Q}\left( \frac{b}{8} +\log\frac{1 + \sqrt{1+\frac{b}{z_{\galrho s}}}}{2} \right)\nonumber\\
  -& \log\frac{b}{4} + \frac{\alpha}{32Q}(b-4)(4z_{\galrho s} +3b +12)\nonumber\\ 
+&\frac{\alpha\galrho}{2Q}\Bigg[ \frac{b}{2}\left[ G(\frac{z_{\galrho s}}{b},0) + \frac{1}{2}\log\left(\frac{b}{z_{\galrho s}} \right) \right]\nonumber\\
+&\frac{\alpha\galrho b}{2Q\sqrt{z_{\galrho s}(z_{\galrho s}+b)}}\bigg[ G(\frac{z_{\galrho s}}{b},0) - G(\frac{z_{\galrho s}}{b},\frac{z_{\galrho s}}{b})\nonumber\\
-&\left( \frac{z_{\galrho s}}{b}-\frac{\sqrt{z_{\galrho s}(z_{\galrho s}+b)}}{b} \right)\log\left(\frac{b}{z_{\galrho s}}\right) \bigg] \Bigg] \nonumber\\
-& \frac{\alpha\galrho}{Q} r
+ \frac{\alpha(1+\galrho)}{Q}\left(s+\log(1-s)\right)
\Bigg].
\end{align}
\end{cor}

\section{Analysis}
\subsection{Dependence of $E(r)$ on $\alpha=T/N$}

\begin{figure}[h]
\centering
  \includegraphics[scale=.85]{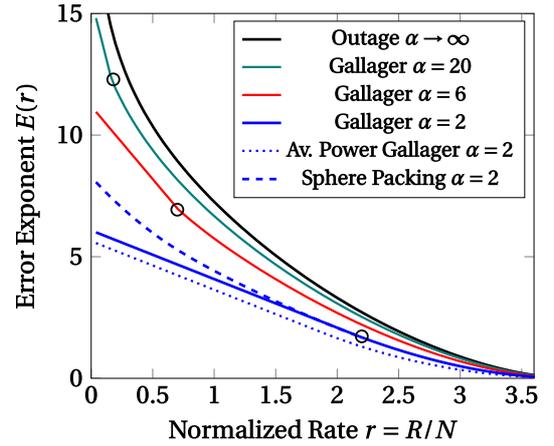}
\caption{The Gallager error exponent $E(r)$. As $\alpha$ is increased, the curves for $E(r)$ approach the outage probability exponent $E_{out}(r)$ \cite{kazakopoulos2011living} (dashed). The small circles indicate the points  where $r=r_1$. For $\alpha=2$ we also depict Gallager exponent for the average power constraint ($s=0$) and the Sphere Packing Bound error exponent (dot-dashed). Parameter values used are: $\beta =3$, $\mbox{SNR} = \sigma^{-2}= 20$, $Q = 1$.}
  \label{Fig:Beta_gen}
  \end{figure}

In Fig.~\ref{Fig:Beta_gen}, we plot the Gallager error exponent for various values of $\alpha$. We see that increasing $\alpha$ brings the error curve closer to the error exponent $E_{out}(r)$ of the infinite codelength outage probability introduced in \cite{kazakopoulos2011living}. This convergence can be seen directly in \eqref{eq:mu_x_2}-\eqref{Eq:r_bar}. As $\alpha\to\infty$, $\galrho\to 0$, so that $\alpha\galrho=O(1)$ and the solution converges to that of \cite{kazakopoulos2011living}. 

It is important to point out here that the assumption that the receiver knows the channel matrix necessitates the existence of some training overhead, which becomes significant when the number of channel uses $T$ becomes comparable to the number of transmit antennas $N$. We do not take into account this issue here, assuming instead that the training takes place through some parallel channel. However, an effective way to incorporate training is to replace $\alpha$ by $\alpha-1$, since it takes roughly $N$ channel uses to train the $N$ transmit antenna channels.   

\subsection{$r \leq  r_1$ and Comparison with Sphere Packing Bound}
 The circles in Fig.~\ref{Fig:Beta_gen} correspond to the values $r=r_1=\bar{r}(\galrho=1)$, below which the Gallager error exponent becomes linear in $r$. This behavior is due to the fact that the value of the error exponent in \eqref{Eq:E(r)_gen} is the result of the maximization with respect to the parameter $\galrho$ over the unit interval $\galrho\in [0,1]$. For $r<r_1$ the maximum lies outaside this interval and hence $\galrho$ remains fixed to unity. Hence the error exponent in \eqref{Eq:E(r)_gen} becomes linear in $r$. Extending the $\galrho$-maximization interval to $\RR^+$ provides the so-called sphere-packing error exponent \cite{gallager1968information}. In  Fig.~\ref{Fig:Beta_gen} we include the sphere-packing exponent in the case of $\alpha=2$ (dash-dot) for comparison. As expected, for rates above the value of  $r=r_1$ indicated by a circle, the error exponent coincides with the Gallager random coding exponent, while for $r<r_1$ (corresponding to solutions with $\galrho>1$) the sphere-packing exponent is higher.

\subsection{Region $r \approx r_{erg}$ and Comparison with \cite{hoydis2015second} }
 
The region close to $r=r_{erg}$ is interesting because the error exponent $E(r)$ vanishes  and hence the error probability is maximal. It is easy to see that $\frac{d E(r)}{d r}=-\alpha\galrho(r) $, where $\galrho (r)$ is the solution of the equation $r = \bar{r}(\galrho)$ in \eqref{Eq:r_bar} for $r>r_1$. From \eqref{Eq:r_bar}, we see that when $\galrho\to 0$, then $r\to r_{erg}$. This implies that $E(r_{erg})=0$ is a global minimum, since,  taking advantage of the convexity of the supremum operation with respect to $\galrho$ and $s$, it can be shown that $E(r)$ is a convex function of $r$ \cite{karadimitrakis2014outage}. Therefore, close to $r = r_{erg}$, we can write
\begin{align}
\galrho(r) = (r-r_{erg})\galrho'(r_{erg})
\end{align} 
where $\galrho'(r) = \frac{d\galrho}{dr}$. 
Let us define $v_{\alpha}$ through 
\begin{align}
\galrho'(r_{erg}) = -\frac{1}{\alpha v_{\alpha}}.
\end{align}
 The left-hand-side of the above equation is easy to evaluate since $\frac{d\galrho(r)}{d r}\frac{d\bar{r}(\galrho)}{d\galrho}=1$. Hence, by differentiating $\bar{r}(\galrho)$ and expressing its value at ${\galrho =0}$, we obtain
  \begin{align}
  \label{Eq:Gauss_approx}
  E(r) = \frac{(r-r_{erg})^2}{2 v_{\alpha}} + o\left( (r-r_{erg})^2 \right).
  \end{align}
In the above, $v_{\alpha}$ can be expressed as
\begin{align}\label{eq:valpha}
v_{\alpha} = v_{\infty} + \frac{\delta v }{\alpha},
\end{align}
where $v_{\infty}$ is the infinite codelength dispersion given in \eqref{Eq:vopt_def} and $\delta v>0$ has the simple form
\begin{align}\label{eq:deltav}
\delta v = 2 g_0 - g_0^2,
\end{align}
where $g_0$ is given by 
\begin{align}\label{eq:deltav1}
g_0 = \int_{a_0}^{b_0}  \frac{xp_0(x)}{x+\s2} dx =\frac{\left(\sqrt{\s2+b_0}-\sqrt{\s2+a_0}\right)^2}{4},
\end{align}
where $p_0(x)$ is the Marcenko-Pastur distribution given in \eqref{Eq:MarcPastur} and $a_0,~b_0$ its endpoints. It is worth pointing out that the last term in \eqref{eq:deltav} is the correction due to the peak-power codeword constraint \eqref{Eq:Energy_Contraint}. We see that the Gallager error exponent $E(r)$, which is valid for all rates $r< r_{erg}$ takes a quadratic form akin to the exponent of a normal distribution for rates close to $r_{erg}$. This is analogous to the case of infinite codelengths discussed in Section \ref{sec:channel_model_cap}. 
\eqref{Eq:Gauss_approx} is valid when $|r_{erg}-r|\ll 1$, in order for the error exponent to be small. However, it is also implicitly assumed that $N|r_{erg}-r|\gg 1$, so that the term $N^2 E(r)$ in the error probability exponent (see \eqref{Eq:EN_def}) is the dominant one. Hence this is exactly the moderate deviations regime discussed for general single link systems  in \cite{altug2014moderate}.  An important point that can be drawn from the form of \eqref{eq:deltav1} is that it depends only on the empirical distribution of eigenvalues, which in this case happens to be the Marcenko-Pastur distribution. Therefore, $v_\alpha$ can be calculated for other channel models for which $v_\infty$ and $p_0(x)$ are known.

In \cite{hoydis2015second}, the authors obtained bounds on the optimum average probability of error for MIMO systems when the normalized rate of the code $r=R/N$ approaches the ergodic rate $r_{erg}$ such that $N|r-r_{erg}|=O(1)$ in the limit that $N,K,T$ become large with fixed ratios. In this limit, they show that the error probability is bounded between two Gaussian distributions with variances (or dispersions) given (in their notation) by $\theta _{-}$, which can be expressed as
\begin{align}
\theta_{-} = \alpha v_\infty + \frac{1}{2}\left(\beta+1-\frac{\s2(\beta+1)+(\beta-1)^2}{\sqrt{(\s2+a_0)(\s2+b_0)}}\right),
\end{align}
and $\theta _{+}=\alpha v_\alpha$, respectively. Therefore, the Gallager random coding exponent with Gaussian input saturates the upper bound in the dispersion derived by \cite{hoydis2015second}.

\begin{figure}[h]
\centering
\includegraphics[scale=.85]{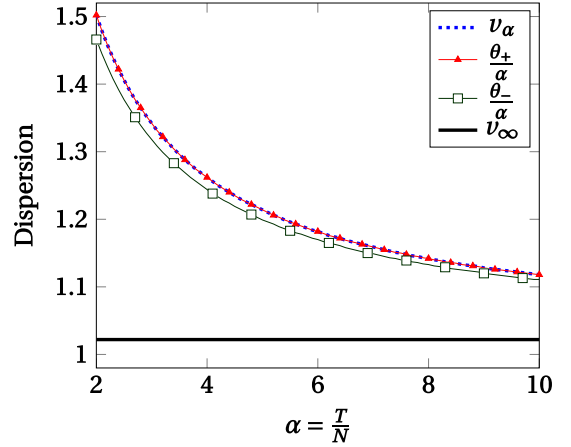}
\caption{The dispersion at the Gaussian limit ($\galrho =0$) using the asymptotic method and the method of induced ergodicity of \cite{hoydis2015second}. The $v_\alpha$ and $\frac{\theta_+}{\alpha}$ curves are identical; $\mbox{SNR} = \sigma^{-2}= 20$, $\beta =3$, $Q = 1$.}
\label{Fig:Gauss_approx_beta_1}
  \end{figure}

\subsection{Impact of Fading}

\begin{figure}[h]
\centering
\includegraphics[scale=.8]{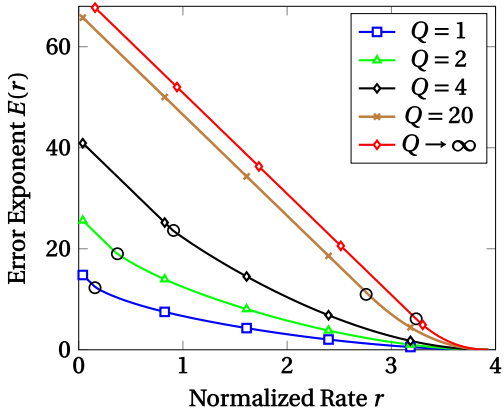}
\caption{The error exponent for the Gallager bound with power constraint at the limit of $N \rightarrow \infty$; $\beta =3$, $\mbox{SNR} = \sigma^{-2}= 20$, $\alpha =20$.The small circles indicate the points of behavior change ($r=r_1$).}
\label{Fig:E_Q}
  \end{figure}

The case $Q>1$ models the realistic situation where the channel varies during the transmission of the codeword. Specifically, the channel matrix ${\bf H}$ changes ($Q$ times) during the codelength $T$. It is assumed here that the receiver knows each channel realization, either using an additional pilot signal or by using part of the codeword as pilot (in which case $T$ will represent the data-transmitting part of the codeword). In Fig.~\ref{Fig:E_Q} we can see the behavior of the error exponent for increasing values of $Q$. As $Q$, the number of independent fading blocks within a codeword increases, the error exponent $E(r)$ also increases, signifying lower error probabilities. To understand the behavior for large $Q$, we prove the following result.  

\begin{theorem}[$Q\rightarrow \infty$ Limit of $E(r)$]
\label{Th:Q_infty}
\begin{align}
\label{Eq:E_Q_infty}
\lim _{Q\rightarrow \infty}E(r)=&\alpha\max_{\substack{\galrho \in [0,1]\\ s\in [0,1)}}\bigg[\galrho r_{erg}(\beta,z_{\galrho s}^{-1})- \galrho r\nonumber\\
+&(1+\galrho)(s +\log(1-s))\bigg].
\end{align}
For fixed $\galrho$, the maximum over $s$ in the above equation is attained at the value
\begin{align}\label{eq:s_condition0}
  s = \frac{\galrho}{4(1+\galrho)} \left(\sqrt{z_{\galrho s}+b_0}-\sqrt{z_{\galrho s}+a_0}\right)^2.
\end{align}
Defining the function 
\begin{align}
\bar{r}(\galrho)=\log(1-s)+r_{erg}\left(\beta, z_{\galrho s}^{-1}\right),
\end{align}
and setting $r_1=\bar{r}(1)$ we have 
\begin{align}
\galrho(r) = \left\{
\begin{tabular}{ c c }
 $1$ & $r\leq r_1$\\ 
 ${\bar r}^{-1}(r)$ & $r>r_1$     
\end{tabular}
\right.
\end{align}
where $\bar{r}^{-1}$ indicates the inverse function of $\bar{r}$. 
\end{theorem}
From the above theorem we conclude that for fast-fading, and therefore large values of $Q$ it is the ergodic rate that determines the behavior of the error exponent. When $r\approx r_{erg}$ we can once again expand $E(r)$ in powers of $r-r_{erg}$ to obtain
\begin{align}
  \label{Eq:Gauss_approxQinf}
  E(r) = \frac{(r-r_{erg})^2}{2 \,\,\delta v} + o\left( (r-r_{erg})^2 \right),
  \end{align}
where $\delta v$ is given in \eqref{eq:deltav}.

\section{Conclusion}
In this paper we have applied random matrix theory to calculate an analytic expression of the Gallager bound for finite codelength for block fading channels with $Q$ independent fading blocks within a codeword. This method is valid for arbitrary normalized rates $r<r_{erg}$, in the large $N,K,T$ limit. As expected, the error exponent increases with $Q$, resulting to a lower error probability. The limit $Q \rightarrow \infty$ is characterized by $r_{erg}$. Furthermore, when the normalized rate $r$ becomes close to $r_{erg}$ the Gallager exponent becomes asymptotically equal to an upper bound of the fixed optimal error analysis derived recently in \cite{hoydis2015second}. The methodology we have used can be generalized to other MIMO channels for which the joint eigenvalue is known. For example, we have recently applied it to obtain the Gallager exponent for fiber-optical MIMO channels \cite{Karadimitrakis2017_GallagerOMIMO}. Other cases for which the joint eigenvalue distribution of the effective channel is known and hence this methodology can be directly applied by using the appropriate weight function $w(x)$ in \eqref{Eq:JPDF}, include the uplink MU-MIMO channel \cite{Forrester2010_book} and the Amplify-and-Forward channel \cite{Chen2013_RMT_WirelessRelaying}. It should be noted that more general Gaussian channels, which do not have a known joint eigenvalue distribution can be analyzed in similar ways using the replica method \cite{moustakas2003mimo}.

\appendices
\section{Proof of Theorem~\ref{Th:E_beta_gr_1}}
\label{Appendix:Proof1}

For $Q=1$ we study the limit
\begin{align}
E(r) =&-\lim _{N\rightarrow \infty} \frac{1}{N^2}\log \mathbb{E}_{\mathbf{H}}\left[ e^{-N^2f[\mu_N]} \right],
\label{Eq:BigN}
\end{align}
where $\mu_N(x) = \frac{1}{N}\sum_i \delta (x-\lambda_i)$, $\lambda_i$ are the eigenvalues of $\mathbf{HH}^\dagger$ and $f[p]$ is defined on $\mathcal{M}(\mathbb{R}^+)\rightarrow\mathbb{R}$, where $\mathcal{M}(\mathbb{R}^+)$ is the space of probability measures on $\mathbb{R}^+$ as
\begin{align}
	f[p] =&
	 \alpha \max_{\substack{\galrho \in [0,1]\\ s\in [0,1)}}\bigg\lbrace\galrho\int_{\mathbb{R}^+}\log\left(1 + \frac{x}{z_{\galrho s}} \right)p(x)dx- \galrho r\nonumber\\
	  +&(1+\galrho)(s +\log(1-s))\bigg\rbrace, 
	 \label{eq:fp_def}
\end{align}
where the argument of $\max$ is defined as $ g[\galrho,s,p]$. It is therefore important to show a number of properties of $f[p]$ and $g[\galrho, s, p]$. First,  when $\galrho=s=0$, the function $g$ vanishes, i.e. $g[0,0,p]=0$, so that $f[p]\geq 0$. Second, $f[p]$ is  continuous in $p$, for which Berge's Maximum theorem \cite{berge1963topological} can be invoked. Third, $f[p]$ is convex in $p$, which can be shown directly from its definition. Fourth, $g[\galrho,s,p]$ is quasi-concave in $\galrho$, $s$. To show this we start by noting that, excluding the term $\galrho s$ in \eqref{eq:fp_def}, $g[\galrho,s,p]$ is concave in both $\galrho$, $s$. Hence, since $\galrho s$ is quasi-concave, so is $g[\galrho,s,p]$. Therefore for all $p\in \mathcal{M}(\mathbb{R}^+)$ for which the integral 
$\left|\int_0^\infty \log(x) p(x) dx\right|<\infty$, $g[\galrho,s,p]$ has a global maximum in $\galrho\in [0,1]$, $s\in [0,1)$. 

\subsection{Varadhan's Lemma}

We now wish to invoke Varadhan's Lemma. To do so, we first provide the following definitions: 
\begin{definition}\cite{dembo2009large} \normalfont
A rate function $I[p]$ is a lower semicontinuous mapping $I:\mathcal{M}(\mathbb{R}^+)\to [0,\infty]$, for which all level sets are closed. If, in addition, the level sets are compact, then $I[p]$ is called a good rate function. 
\end{definition}
\begin{definition}\cite{arous1997large, dembo2009large} \normalfont
The probability law $\mu_N$ satisfies the large deviation principle in the scale $N^2$ with rate function $I$ if, for all subsets of $\Gamma\subset\mathcal{M}(\mathbb{R}^+)$
\begin{align}
-\inf_{p\in\Gamma^o} I[p]&\leq \liminf_{N\to\infty}\frac{1}{N^2}\log \mu_N(\Gamma)
\leq \limsup_{N\to\infty}\frac{1}{N^2}\log \mu_N(\Gamma)\nonumber\\
&\leq -\inf_{p\in\bar{\Gamma}} I[p],
\end{align}
where $\Gamma^o$ and $\bar{\Gamma}$ are the interior and closure of $\Gamma$, respectively.
\end{definition}

We now note that $f[p]$ is continuous. In addition, since $f[p]\geq 0$ for every $p$, then for any $\gamma> 0$
\begin{align}
\limsup_{N\rightarrow\infty}\frac{1}{N^2}\log\mathbb{E}\left[ e^{-\gamma N^2f[\mu_N]} \right] < \infty.
\end{align}
Furthermore, in \cite{HiaiPetz1998_LargeDeviationsWishartEigenvalues} it was shown  that $\mu_N$, the probability law of the $\{\lambda_i\}$ satisfies a large deviation principle with good rate function given by 
\begin{align}
\label{Eq:Good_Rate}
I[p]&= \int\int \log|x-y|p(x) p(y)dxdy \nonumber\\
&+ \int\left( x-(\beta-1)\right)\log(x)p(x)dx\nonumber\\
&-\frac{1}{2}\left( 3\beta -\beta ^2\log \beta +(\beta-1)^2\log(\beta-1)\right).
\end{align}
As a result, Varadhan's Lemma can be applied to \eqref{Eq:BigN} to show that the limit exists and is equal to
\begin{align}
\label{Eq:inf_max}
E(r) = \inf_{p\in\mathcal{M}(\mathbb{R^+})}\left( f[p] + I[p]\right).
\end{align}

Furthermore, it is possible to show that $I[p]$ is a convex function of $p$. This follows directly from \cite{arous1997large, kazakopoulos2011living} by observing the quadratic dependence of $I[p]$ in $p$. Therefore, since $f[p] + I[p]$ is convex in $p$ its infimum has a unique solution. Taking into account the definition of $f[p]$ and its concave-convex properties discussed above, we may apply Sion's theorem \cite{sion1958general} to exchange the order in which the $\max-\inf$ are applied. Therefore, $E(r)$ in \eqref{Eq:inf_max} can be expressed as
\begin{align}
\label{Eq:Max_Inf_I}
E(r) = \max_{\substack{\galrho\in [0,1]\\s\in[0,1)} }\inf_{p\in\mathcal{M}(\mathbb{R^+})}\left( g[\galrho,s,p] + I[p]\right).
\end{align}

\subsection{Explicit Solution of optimum $p(x)$ and Evaluation of $E(r)$}
To solve the above optimization problem \eqref{Eq:Max_Inf_I}, we  introduce the Lagrangian functions
\begin{align}
\label{Eq:L0}
\mathcal{L}_0[p,c] = f[p] + I[p] -c\left(\int p(x)dx -1 \right),
\end{align}
\begin{align}
\label{Eq:L1}
&\mathcal{L}_1[p,c,\galrho,s] = \mathcal{L}_0[p,c]+\alpha(\galrho+1)\left(s+\log(1-s)\right)\\
& + \alpha\galrho\left( \int\log \left(1+\frac{x}{z_{\galrho s}} \right)p(x)dx -r  \right).\nonumber
\end{align}
Since $\mathcal{L}_1$ is convex in $p$ and concave in $c$, $\galrho$ and $s$, the saddle point is unique \cite{boyd2004convex} and we obtain
\begin{align}
\label{Eq:E(r)_sup}
E(r) = \sup_{c,\galrho, s}\inf_p\mathcal{L}_1[p,c,\galrho,s].
\end{align}

Taking advantage of the convexity in $p$, in order to find the infimum of $\mathcal{L}_1$ we will take the functional derivative with respect to $p$, which is defined as
\begin{align}
\label{Eq:Perturb}
\delta \mathcal{L}_1[p] = \frac{d}{dt}\bigg|_{t=0}\mathcal{L}_1[p^* +t\phi],
\end{align}
where $(p+ t\phi) \in \mathcal{M}(\mathbb{R}^+)$ and $\phi$ is a test function.
This can be re-written as 
\begin{align}
\delta \mathcal{L}_1[p] =\int\phi(x)\Psi[p^*,x]dx,
\label{Eq:Perturb2}
\end{align}
where
\begin{align}
\label{Eq:Psi}
\Psi[p^*,x] =& -2\int p(y)\log|x-y|dy\\
 +&x -(\beta -1)\log(x)
 -c +\alpha\galrho\log\left( 1 + \frac{x}{z_{\galrho s}}\right).\nonumber
\end{align}
At the minimum, \eqref{Eq:Perturb2} must vanish identically for all $\varphi$, thus $\Psi[p^*,x] = 0$ and it follows that
\begin{align}
2\int \log|x-y|p^*(y)dy =& x-(\beta-1)\log(x)\nonumber\\
 -&c + \alpha\galrho\log\left( 1 + \frac{x}{z_{\galrho s}}\right).
\end{align} 
Next, we differentiate \eqref{Eq:Psi} with respect to $x$ to obtain
\begin{eqnarray}
 2 \text{PV}\int \frac{p^*(x)}{x-y}dx = 1 - \frac{\beta-1}{x} + \frac{\alpha\galrho }{z_{\galrho s}+x}, 
 \label{Eq:PV}
\end{eqnarray}
where PV denotes the principle value. 
Following \cite{tricomi1957integral}, the solution of the last equation is given by
\begin{align}
\label{eq:mu_x_full}
  p^*(x) &= \frac{1}{2\pi\sqrt{(x-a)(b-x)}}\\
   \times&\left[-x-(\beta-1)\frac{\sqrt{ab}}{x}+ \frac{\alpha\galrho\sqrt{(z_{\galrho s}+a)(z_{\galrho s}+b)}}{(x+z_{\galrho s})} + C\right],\nonumber
\end{align}
where $C$ is an unknown constant and $a,~b$ are the (unspecified) endpoints of the support of $p(x)$. Since $p^*(x)$, if it exists, is unique, we search for a solution among continuous, non-negative, normalized functions over $x\in (0,\infty)$. Continuity at $x=b$ demands that $p^*(b) = 0$, which fixes the value of $C$ above, while continuity at $x =a$, i.e. $p^*(a) = 0$ results to  \eqref{eq:mu(a)=0}. Furthermore, the normalization condition on $p(x)$, $\int_a^b p(x)dx = 1,$ 
results to \eqref{eq:mu_norm_condition}. In Appendix~\ref{Appendix:Unique_a_b} it is shown that for fixed $\galrho$ there is a unique solution of \eqref{eq:mu(a)=0} and \eqref{eq:mu_norm_condition} for $0<a<b$. For $s \in [0, 1)$, $\mathcal{L}_1$ is concave. Hence the maximum over $s$ results when the first derivative of $\mathcal{L}_1[p^*]$ with respect to $s$ vanishes, hence \eqref{eq:s_condition}.

Once we have determined the value of $s$ as a function of $\galrho$, we now search for the optimal value of $\galrho$. Extending its support to  $\galrho \in [0, \infty)$, the extremal value of $\galrho$ is determined by \eqref{Eq:r_bar}. However, since the optimization of $\galrho$ is over $[0,1]$, then there are two possible types of solution: For  $r_{erg}\geq r \geq r_1$, the optimal value of $\galrho<1$, hence the value of $\galrho$ is determined by \eqref{Eq:r_bar}. In contrast, for $r<r_1=\bar{r}(1)$, the optimal value of $\galrho$ is fixed to the boundary of the region, i.e. $\galrho=1$.

Having determined the values of $a$, $b$, $s$, $\galrho$, we may now integrate the expression in \eqref{Eq:L1} to evaluate $E(r)$. The integrals appear in \eqref{eq:mu_x_2} in Theorem 1. All single integrals over $p(x)$ can be evaluated in closed-form directly. \eqref{Eq:Psi} can be used to simplify the double integral over $p$ into a single one, which then can be evaluated directly. The value of $c$ in \eqref{Eq:Psi} can be obtained by evaluating $\Psi[p,x]$ at $x=a$. Finally, to obtain the value of $E(r)$ for general $Q$, we make the substitution $\alpha\to\frac{\alpha}{Q}$ and $E(r)\to Q E(r)$ as discussed in Remark \ref{rem:E(r,Q)}.


\section{Proof of uniqueness of solution of \eqref{eq:mu(a)=0},\eqref{eq:mu_norm_condition} }
\label{Appendix:Unique_a_b}
To show that \eqref{eq:mu(a)=0} and \eqref{eq:mu_norm_condition} have a unique solution, we observe that the normalization integral $n(b)= \int_{a(b)}^bp(x)dx$ is an increasing function of $b$
since its derivative can be expressed as
\begin{align}
n'(b) = \frac{1}{4z_{\galrho s}}\left( 1+ \frac{(\beta-1)z_{\galrho s}}{\sqrt{a(b)b^3}}\right)\frac{z_{\galrho s}-a(b)}{z_{\galrho s}+b} >0.
\end{align}
As $a(b)$ is a decreasing function and bounded below by $0$, we have $\lim_{b\rightarrow \infty} n(b) = +\infty$ and by continuity there will be a unique $b^*$ such that $n(b^*) =1$. Thus, both $a^* = a(b^*)$ and $b=b^*$ will be the unique solution to \eqref{eq:mu(a)=0} and \eqref{eq:mu_norm_condition}.

\section{Proof of Theorem \ref{Th:Q_infty}}

Let us re--write \eqref{Eq:inf_max} as:
\begin{align}
\label{Eq:J_Q}
E(r) = Q\underbrace{\left( \frac{1}{Q}f[p^*] + I[p^*]\right)}_{J_Q[p^*]},
\end{align} 
where $p^*$ is the function $p$ at the infimum of $J_Q[p]$. To examine the behavior of the error exponent $E(r)$ for large $Q$, we analyze the derivative of $ J_Q$ with respect to $Q$. Since $J_Q[p]$ is stationary at $p^*$, its variations with respect to $p$ vanish at $p^*$. Hence, since both $f[p^*]$ and $I[p^*]$ do not depend explicitly on $Q$ we obtain
\begin{align}
\label{Eq:dJ/dQ}
\frac{dJ_Q}{dQ}= \frac{\partial J_Q}{ \partial Q} = -\frac{1}{Q^2}f[p^*].
\end{align}
Now, as $Q$ grows,  it can be seen from \eqref{eq:mu(a)=0} and \eqref{eq:mu_norm_condition} that $p^*(x)$ converges to the Mar\v{c}enko-Pastur distribution and $a$, $b$ converge to $a_0$ and $b_0$ respectively. Hence, $f[p^*]$ becomes $f[p_0]$. Therefore if we integrate \eqref{Eq:dJ/dQ} between $(Q,\infty)$ we find that to leading order in $Q$,  $J_Q \approx \frac{f[p_0]}{Q}$. Multiplying $J_Q[p^*]$ with $Q$ as in \eqref{Eq:J_Q}, we obtain \eqref{Eq:E_Q_infty}.

%
%
%
%

\ifCLASSOPTIONcaptionsoff
  \newpage
\fi



\bibliographystyle{IEEEtran}

\begin{IEEEbiographynophoto}{Apostolos Karadimitrakis}
obtained his B.Sc. degree in Physics from National and Kapodistrian University of Athens (NKUA), an M.Eng in Electronics and Radio-communications from Department of Physics (NKUA) in 2008. He worked at the private sector. He is currently pursuing his PhD degree.
\end{IEEEbiographynophoto}

\begin{IEEEbiographynophoto}{Aris L. Moustakas}  (SM'04)
holds a B.S. degree in physics from Caltech  and M.S. and Ph.D. degrees in theoretical condensed matter physics from Harvard University, respectively.
He joined Bell Labs, Lucent Technologies, NJ in 1998, first in the Physical Sciences Division and then also in the Wireless Advanced Technology Laboratory. He is currently a faculty member at the Physics Dept. of the National Kapodistrian University of Athens, Greece. During 2013-2014 he held a DIGITEO Senior Chair in Orsay, France.
Dr. Moustakas served as Associate Editor for the IEEE TRANSACTIONS ON INFORMATION THEORY between 2009-2012. His main research interests lie in the areas of multiple antenna systems and the applications of game theory and statistical physics to  communications and networks.
\end{IEEEbiographynophoto}

\begin{IEEEbiographynophoto}{Romain Couillet}  
received his MSc in Mobile Communications at the Eurecom Institute and his MSc in Communication Systems in Telecom ParisTech, France in 2007. From 2007 to 2010, he worked with ST-Ericsson as an Algorithm Development Engineer on the Long Term Evolution Advanced project, where he prepared his PhD with Supelec, France, which he graduated in November 2010. He is currently a full professor in the LSS laboratory at CentraleSupélec, France. His research topics are in random matrix theory applied to statistics, machine learning, signal processing, and wireless communications. He is an IEEE Senior Member. In 2015, he received the HDR title from University ParisSud. He is the recipient of the 2013 CNRS Bronze Medal in the section "science of information and its interactions", of the 2013 IEEE ComSoc Outstanding Young Researcher Award (EMEA Region), of the 2011 EEA/GdR ISIS/GRETSI best PhD thesis award, and of the Valuetools 2008 best student paper award.
\end{IEEEbiographynophoto}

\vfill

\end{document}